\documentclass[12pt,svgnames]{article}
\usepackage[style=apa,natbib=true,language=english,doi=false]{biblatex}
\bibliography{arroyo-arenhart-THEORIA-2024.bib}
\usepackage{csquotes,amsmath,epigraph,forest,mathrsfs,ebgaramond-maths}
\usepackage[english]{babel}
\usepackage{authblk}

\usepackage[colorlinks,allcolors=NavyBlue]{hyperref}

\title{\bf Quantum ontology de-naturalized: What we \textit{can't} learn from quantum mechanics}

\author[1,2,4]{Raoni Arroyo\thanks{Corresponding author.}}
\affil[1]{Department of Philosophy, Communication, and Performing Arts, Roma Tre University, Rome, Italy, raoniarroyo@gmail.com}
\affil[2]{Centre for Logic, Epistemology and the History of Science, University of Campinas}

\author[3,4]{Jonas R. Becker Arenhart}
\affil[3]{Federal University of Santa Catarina, Department of Philosophy, Florianópolis, Brazil, jonas.becker2@gmail.com}
\affil[4]{Research Group in Logic and Foundations of Science (CNPq)}

\date{Forthcoming in \textit{THEORIA}, Special Issue: ``Quantum Mechanics and Reality'', 2024}

\begin{document}
\sloppy\raggedbottom
\maketitle

\begin{abstract}
Philosophers of science commonly connect ontology and science, stating that these disciplines maintain a two-way relationship: on the one hand, we can extract ontology from scientific theories; on the other hand, ontology provides the realistic content of our scientific theories. In this article, we will critically examine the process of naturalizing ontology, i.e., confining the work of ontologists merely to the task of pointing out which entities certain theories commit themselves to. We will use non-relativistic quantum mechanics as a case study. We begin by distinguishing two roles for ontology: the first would be characterized by cataloging existing entities according to quantum mechanics; the second would be characterized by establishing more general ontological categories in which existing entities must be classified. We argue that only the first step is available for a naturalistic approach; the second step not being open for determination or anchoring in science. Finally, we also argue that metaphysics is still a step beyond ontology, not contained in either of the two tasks of ontology, being thus even farther from science.

\end{abstract}

\paragraph{Short summary:} Within the relationship between ontology and science, it is argued that the prospects of a naturalistic ontology are restricted to existence questions, and that broader ontological classification---as well as metaphysical theorizing---is beyond scientific determination. Non-relativistic quantum mechanics is used as a case study.

\epigraph{``What we call science has the sole purpose of determining what \textit{is}.''}{\citet{einstein1951}}

\section{Introduction}\label{intro}

After Quine, it seems commonplace to say that ontology deals with \textit{existence} questions. Philosophers of science commonly connect ontology and science by stating that the disciplines maintain a two-way relationship: on the one hand, we can read off at least some of the existential commitments of our best scientific theories; on the other, advancing an ontology gives realistic content to a scientific theory, i.e., it says what the theory in question is about, whether observable or not. The project for \emph{naturalizing ontology} maintains, roughly speaking, that science should guide ontology, thus mainly confining the work of ontologists to the task of exhibiting a catalog of entities that exist according to our best theories. In this paper, we present what is involved in this kind of project and evaluate its scope and some of its limits. 

We are taking seriously the idea that our best science does provide for a furniture of the world. This characterization seems so minimal, that realists and empiricists agree on it. Think of van Fraassen's \citeyearpar[242]{vanfraassen1991} ``question of interpretation'', and Ruetsche's \citeyearpar[3433]{ruetsche2015shaky} ``realist content'', viz., ``\textelp{} an account of what the worlds possible according to [a given scientific theory] are like''. A peaceful point in this discussion is that a scientific theory has an unobservable ontology (interpreted literally). Apart from instrumentalists, who are out of our scope here, no one denies that there has to be an ontology. The problem is how to obtain it, and what should one do with it, once obtained; i.e. should one adopt a realist stance? Or better be an empiricist/agnostic about unobservable entities in the furniture? Although this is one of the greatest problems in the philosophy of science, nothing of what follows for the major points of this paper hangs on the latter kind of problem.\footnote{It should be clear that empiricists in general don't deny that the theory makes ontological/existential postulates. This is denied by instrumentalists. What is denied by empiricists is that we should be \textit{committed} to such posits. The empiricist stance is characterized, among other things, by resisting ``\textelp{} the temptation to reify what is posited in one's ontology'' \citep[8]{bueno2019}. The difference is not in the language and in recognizing the posits of the theories, but in the type of commitments that are adopted \textit{about} them.}

From here, then, we seem to have at least two options: either conceive that the ontology comes with the theory, or else that it comes from an interpretative addition to the theory. That is the problem of furnishing identity conditions for theories and for interpretations; it is a huge topic, and cannot be dealt with here.\footnote{~We will not discuss here whether the term ``interpretation'' is a good one for quantum foundations; the interested reader may find helpful insights about such a debate in \citet{Maudlin1995measurementproblem}, \citet{cirkovic2005physics}, \citet{Ruetsche2018GetRealQM} and \citet{durrlazarovici2020understandingqm}. In the foregoing, we'll straightforwardly call the solutions for the measurement problem interchangeably ``quantum theories'' and ``quantum interpretations''. Nothing that follows strictly depends upon such a distinction. The interested reader might find a discussion on this topic in \citet{arenhart-arroyo2023instante}.}
In any of these cases, quantum mechanics has an ontological content, and this is independent of having a realist or empiricist stance (except an instrumentalist one, which is not in our scope). Science is expected to tell us how the world is, or how the world could be. In both cases, ontology---in the sense we are using this term here, i.e. as concerned exclusively with existence questions---does come from science. Chakravartty makes this clear in the following.

\begin{quote}
    In making the sorts of judgments that are required in order to bring experience to bear in constructing scientific knowledge of the observable world---in distinguishing veridical from non-veridical experience, in determining the quality of observations, and in describing and interpreting it all---it is inevitable that we classify that which is experienced (various objects, events, processes, and properties) into categories for descriptive purposes. This all by itself is a banal truth: it would be impossible to make any of the judgments just enumerated unless we are willing and able to ``carve'' the world into observable categories of things such that we can communicate with one another about these things and thereby coordinate scientific and other activity. In other words, we must taxonomize the content of experience so that it may function as empirical evidence and knowledge. \citep[204]{chakravartty2017prin}\footnote{~Chakravartty's article was originally written in English, but the original text wasn't published. His article was first published in Portuguese. One of us (Raoni Arroyo, together with Ivan F. da Cunha), however, has undertaken the translation into Portuguese, so we have access to the original English text. The above passage, then, quotes the text as originally written but not as originally published.}
\end{quote}

There is a huge amount of literature discussing the intimate relationships between ontology and science.\footnote{~As non-exhaustive list, see \citet{Quine1948-QUIOWT-7}, \citet{Carnap1950-CARESA}, \citet{maudlin2007metaphysics, maudlin2019quantumphil}, \citet{ruetsche2015shaky, Ruetsche2018GetRealQM,}, \citet{esfeld2019measurement}, \citet{durrlazarovici2020understandingqm}.} This article focuses on the metaontological aspects of such a relation, but our focus is not on theory choice in ontology. Rather, we investigate to what extent we can ``extract'' ontological lessons from quantum mechanics; in particular, we focus on the very idea that ontology can be naturalized. Is there a method for extracting ontology from science? There is a tendency in the metaphysics of science that the only part of metaphysics that matters is ontology, which is about what exists, and that only ``naturalized'' ontology, that is, an ontology that is continuous with science (more precisely: \textit{extracted from science}), can have some epistemic and intellectual respectability \citep[see][]{Ladyman2007evmustgo, maddy2007secondphil, arroyo-arenhart2022-synt,maudlin2007metaphysics}. 

The term ``naturalism'' is often poorly defined, as \citet{bryant2020naturalisms} pointed out, with different authors making different uses of the term. For instance, we can find disagreement between Ladyman and Ross' ``metaphysics naturalized'' \citep{Ladyman2007evmustgo} and the Second Philosopher's perspective  \citep{maddy2007secondphil}---to mention two prominent ``brands'' of ``naturalized'' or ``scientific'' metaphysics. According to \citet{Ladyman2007evmustgo}, metaphysics should be continuous with fundamental physics, while \citet{maddy2007secondphil} holds that physics has no privilege over other scientific disciplines \citep[see the discussion in][pp.~222--223 fn.~9.]{maddy2022}. Thus, one finds the literature orbiting around the \textit{continuity} between metaphysics and science, and this notably can be itself spelled out in different---and incompatible---ways:

\begin{quote}
    It is not uncommon to hear that continuity in this context is evidenced by the fact that naturalized metaphysics is `derived from', `based on', or otherwise `inspired' or `motivated' or `constrained by' our best science, which thereby serves as the proper `ground' for metaphysical theorizing. \citep[40--41]{chakravartty2013}
\end{quote}

As each of these goals behind the notion of ``continuity'' isn't equivalent, then one ought to deliver different things depending on the goal. From here on, we'll set the naturalist's goal as to the first one mentioned by \citet[40]{chakravartty2013}, viz., that the goal of naturalized/scientific metaphysics is to \textit{derive} the metaphysics from science. This is a fairly strict sense of metaphysical naturalism; however, as we argued elsewhere  \citep{arroyo-arenhart2022-synt}, the other goals mentioned by \citet[40]{chakravartty2013}---viz., to be ``{`based on', or otherwise `inspired' or `motivated' or `constrained by' our best science}''---do not confer any \textit{distinctively superior} epistemic warrants to naturalized metaphysics in comparison with analytic metaphysics. Here we argue that maybe the naturalized metaphysics qua metaphysics derived from science could have such a privileged fore in comparison with other kinds of non-naturalistic metaphysics. As we understand, this radical sense of naturalized metaphysics is the process of a  ``\textelp{} mechanical extractions of (allegedly) metaphysical claims from scientific theories'' \citep[59]{morganti2015}. So here is how we'll understand from now on what the aim of the project of naturalist metaphysics is:

\begin{quote}
    In scientific metaphysics, it is not enough for a metaphysical proposition to be compatible with current science. It must be endogenous to the empirical enterprise of a given science (generally physics), i.e., it must be derived from that science. \citep[1859]{guaypradeu2020}
\end{quote}

In this paper, we argue that such a goal can only be achieved with a---rather limited---part of ontology (which we don't take itself to be distinctively metaphysics). So there's this twist which aligns us with both radical and moderate naturalists: with radical naturalists, we agree that there's a part of metaphysics that can be extracted/derived from science, namely, part of the ontology; with moderate naturalists, we agree that distinctively metaphysical questions are not subject to this process of extraction/derivation. In order to make this point clearer, we structure the paper as follows. In section \ref{paper:ont:sec-2}, where we narrow our scope to the relation of \textit{ontology} and quantum mechanics, we investigate the traditional approach to the theory's ontology through solutions to the measurement problem. We discuss how quantum mechanics provides, at most, the \textit{catalog} of the theory's entities---its ontological commitments---but does not say anything about the \textit{ontological types} of such entities. We will present the discussion of the ``catalog'' associated with interpretations of quantum mechanics as a showcase in section \ref{paper:ont:sec-3}. In section \ref{paper:ont:sec-4}, we distinguish between metaphysics, ontology, and the two tasks of ontology, respectively: discussing the nature of entities; elucidating the catalog of what exists; and classifying that catalog into ontological categories. We will not deal with metaphysical (qua metaphysics) questions. Section \ref{paper:ont:sec-5} concludes, taking stock of what has been discussed.

\section{The measurement problem}\label{paper:ont:sec-2}

The measurement problem is one of the central issues of QM. There are indeed authors, such as \citet[104]{gibbins1987particles}, who consider the measurement problem to be \textit{the} central problem of QM. After all, one might ask why the philosophy of quantum mechanics is so interested in it.

We seem to have a problem with some basic assumptions made \textit{about} quantum mechanics. In the etymological sense of the term, it is a \textit{paradox}, i.e., something beyond the accepted opinion in general. It is basically a problem with premises that are considered reasonable, which jointly lead to an unacceptable conclusion so that at least one of them must be rejected. To show which are the assumptions whose conjunction is problematic, we will follow Maudlin's (\citeyear{Maudlin1995measurementproblem}) taxonomy, for it became the standard way of stating the measurement problem in QM. Maudlin distinguishes three instances of the measurement problem; for the purposes of this work, the \textit{first} measurement problem is the relevant one, and it is defined as the conjunction of three assumptions:

\begin{quote}
1.A The wave-function of a system is complete, i.e. the wave-function specifies (directly or indirectly) all of the physical properties of a system.\newline 1.B The wave-function always evolves in accord with a linear dynamical equation (e.g. the Schrödinger equation).\newline 1.C Measurements of, e.g., the spin of an electron always (or at least usually) have determinate outcomes, i.e., at the end of the measurement the measuring device is either in a state which indicates spin up (and not down) or spin down (and not up). \citep[7]{Maudlin1995measurementproblem}
\end{quote}

\citet[pp.~7--8]{Maudlin1995measurementproblem} claims that the conjunction $1.\text{A}\land 1.\text{B}\land 1.\text{C}$ is one aspect of the measurement problem, which he calls the ``problem of outcomes''. Strictly speaking, Maudlin's trilemma is not the most precise way of stating the measurement problem. Such a way of enunciating the measurement problem, however, will suffice for the classification that follows.\footnote{~Notably, \citet[10]{Maudlin1995measurementproblem} states that ``[t]he three propositions in the problem of outcomes are not, \textit{strictu sensu}, incompatible'' although he mentions their ``inconsistency'' throughout his paper. This is why \citet[235]{muller2023} claims that Maudlin ``buried'' his \textit{problem of outcomes}. A much more precise account of this aspect of the measurement problem can be found in \citet[230--234]{muller2023}, where it is called ``Reality Problem of Measurement Outcomes'', and presented as a \textit{polylemma} instead of a trilemma, and a rigorous proof of inconsistency is offered.} In order to visualize it, consider Schrödinger's cat ``humane'' scenario:\footnote{~The terminology is due to \citet{putnam2012}, but the analogy that follows is borrowed from \citet[256--260]{carroll2019}. It consists of replacing the flask of poison with a container of sleeping gas.} an idealized closed box containing a cat; a container of sleeping gas; a hammer; a radioactive material (an $\alpha$ particle); a Geiger counter. As a property of the $\alpha$ particle in question, the chances that the particle will decay are both 50\% each. If it decays, then: the Geiger counter will detect the $\alpha$ particle's decay; the hammer will be activated and break the container of sleeping gas, which will make the cat sleep. If the $\alpha$ particle does not decay, then none of this will happen, and the cat will remain awake. At the end of one hour, \textit{something} will happen to the $\alpha$ particle, and both of the cases above are possible subsequent chains of events that follow the state of the $\alpha$ particle.

A typical quantum-mechanical description of physical reality says that physical systems are described by states $|\psi\rangle$ in a Hilbert space $\mathcal{H}$, whose evolution through time is described by a Hamiltonian H in the Schrödinger equation. Let us assume that $|\psi\rangle$ is a quantum-mechanical description for all that is going on in the box of Schrödinger's cat. The Hilbert space is factorized in subspaces for each component: the quantum system $S$, the macroscopic measurement apparatus $A$, and everything else in the environment $E$ in the form $\mathcal{H}=\mathcal{H}_S\otimes\mathcal{H}_A\otimes\mathcal{H}_E$. After one hour, three stages of measurement take place. The pre-measurement state is described as:

\begin{equation}
|\psi\rangle=a|+\rangle_S + b|-\rangle_S\otimes |0\rangle_A\otimes |0\rangle_E
\end{equation}

\noindent the ready state described in the form of:

\begin{equation}
|\psi\rangle=a\left(|+\rangle_S\otimes|+\rangle_A\right) + b\left(|-\rangle_S\otimes|-\rangle_A\right)\otimes|0\rangle_E
\end{equation}

\noindent and finally the measurement state, described as:

\begin{equation}
|\psi\rangle=a\left(|+\rangle_S\otimes|+\rangle_A\otimes|+\rangle_E\right) + b\left(|-\rangle_S\otimes|-\rangle_A\otimes|-\rangle_E\right)
\label{eq:superposition}
\end{equation}

\noindent where: $a$ and $b$ are probability \textit{amplitudes}; $|+\rangle_S$ represents the $\alpha$ particle \textit{decaying}; $|-\rangle_S$ represents the $\alpha$ particle \textit{not decaying}; $|+\rangle_A$ represents the measurement apparatus registering $|+\rangle_S$, e.g. the active hammer; $|-\rangle_A$ represents the measurement apparatus registering $|-\rangle_S$, e.g. the inactive hammer, and $|0\rangle_A$ represents the measurement apparatus registering nothing, e.g. the reset button; finally, $|+\rangle_E$ represents everything else interacting with $S$ and $A$, e.g. you observing a sleeping cat for the case of $|+\rangle_E$, you observing an awake cat for the case of $|-\rangle_E$, and you observing nothing for the case of $|0\rangle_E$ (the box was still closed). To simplify the notation, we might say that the component $|+\rangle_S\otimes|+\rangle_A\otimes|+\rangle_E$ is represented by $|\psi_1\rangle$, and the component $|-\rangle_S\otimes|-\rangle_A\otimes|-\rangle_E$ is represented by $|\psi_2\rangle$.

Regarding the \textit{informal} proof of the inconsistency between assumptions 1.A, 1.B, and 1.C on Maudlin's presentation of the measurement problem, \citet[223]{esfeld2019measurement}, states that:

\begin{quote}
If the entire system is completely described by the wave function [1.A], and if the wave function always evolves according to the Schrödinger equation [1.B], then, due to the linearity of this wave equation, superpositions and entangled states will, in general, be preserved. Consequently, a measurement of the cat will, in general, not have a determinate outcome [1.C] \textelp{}. \citep[223]{esfeld2019measurement}
\end{quote}

So one of the assumptions (1.A, 1.B, 1.C) must be dropped. Someone who answers ``no'' to question 1.A implicitly says that something is escaping the description of the wave function: it does not, therefore, fully describe the physical states of the system in question. So two systems described by the same wave function can be in two different physical states, as there are more properties of the systems than the wave function describes \citep[see][107]{maudlin2011intimations}. Solutions that maintain that the wave function is not complete postulate hidden variables to supplement that wave function. Employing the taxonomy offered in \citet{arroyo-olegario2021}, let us call them QM$_{hid}$. Examples of such interpretations are interpretations of statistical \textit{ensembles} \citep[see][]{Ballantine1970ensemble}, ``Bohmian mechanics'' \citep{bohm1952suggested, sep-qm-bohm}, and the modal-Hamiltonian interpretation \citep{lombardicastagnino2008modalhamilt, lombardi2019modalhamiltonian}, which denies the completeness of the quantum state albeit in a different way.

Question 1.B concerns the evolution of the wave function over time. Responding negatively to question 1.B implies adding a nonlinear dynamic law in the behavior of the wave function. This additional dynamic law is called ``collapse'' or ``reduction''. We will call QM$_{col}$ solutions of this type. According to a collapse-based quantum theory, the wave function is immediately transformed into one of the possible results in certain situations. The ``standard'' interpretation following \citet{vNeum1955mathematical} and the ``spontaneous reduction'' theory \citep*[see][henceforth cited as ``GRW'']{GRW} are among the examples of this type of solution \citep[see also][]{Jammer1974}.

Finally, question 1.C concerns the results of the physical description through the wave function. For anyone who claims that the wave function is complete and denies that there are collapses, there is only one alternative: to deny that there are unique results for a measurement. In such theories, measurements of a system in a superposition of other states, as $|\psi\rangle= c_i|\psi_1\rangle+ c_j|\psi_2\rangle$, do not result in the system being definitively in just one of the states, but something beyond. An experiment starts with just one wave function ($|\psi\rangle$) describing one system to be measured, but the measurement ends up describing two different situations: one resulting in $|\psi_1\rangle$ and another resulting in $|\psi_2\rangle$. Solutions of this type are called ``relative state'' or ``branching-type'', or even ``many-worlds'', and we will call them QM$_{bra}$, among which we can highlight the Everettian quantum mechanics \citep[see][]{everett1957relative,wallace2012emergent}.

We organized the answers in the form of a diagonal in table \ref{tab:maudlin}, which gives us grounds to entertain which interpretation follows by the negation of one of these three simple assumptions.

\begin{table}[ht]
\centering
\caption{Interpretations: the ``no'' diagonal}
\label{tab:maudlin}
\begin{tabular}{cccc} 
1.A & \textbf{N} & Y          & Y           \\
1.B & Y          & \textbf{N} & Y           \\
1.C & Y          & Y          & \textbf{N}  \\ 
\cline{2-4}
\textbf{}    & QM$_{hid}$   & QM$_{col}$   & QM$_{bra}$   
\end{tabular}
\end{table}

\section{On what there is in QM}\label{paper:ont:sec-3}

Formulated in the form of the above trilemma, the current state of the art in the foundations of quantum mechanics has led us to QM$_{hid}$, QM$_{col}$ and QM$_{bra}$ as consistent solutions. This, however, is not enough for many. That is, for those who wish to extract a coherent image of the world according to quantum mechanics, one will have to investigate the ontology of these theories (and what are one's attitudes regarding such ontology, remember, will depend on whether one is a realist or an empiricists). As simply stated by \citet[2]{durrlazarovici2020understandingqm}, ``[t]he ontology of a physical theory specifies what the theory is about''. One way to investigate this issue is to look for the theories' ontological commitments à la \citet[65]{quine1951commitment}, i.e., to check ``\textelp{} what, according to that theory, there is''.
It is important to emphasize that Quine's ontological commitment is fairly neutral with regard to which ontology should be chosen:

\begin{quote}
    It does not dictate any ontology. Quine's criterion is a metaontological tool, providing a strategy for determining what is assumed to exist by candidates for an overall best theory. Rather than recommend some particular ontology, it is designed to compare and adjudicate between rival theories. \citep[252]{janssen-lauret2017}
\end{quote}

Ontology is not, furthermore, something of exclusive concern for philosophers, but it is also crucial for scientific development itself. To illustrate the need for a scientific ontology, Esfeld writes:

\begin{quote}
    \textelp{} an algorithm to calculate measurement outcome statistics is not a physical theory. Physics is about nature, \textit{physis} in ancient Greek. Consequently, a physical theory has to (i) spell out an ontology of what there is in nature according to the theory, (ii) provide a dynamics for the elements of the ontology and (iii) deduce measurement outcome statistics from the ontology and dynamics by treating measurement interactions within the ontology and dynamics; in order to do so, the ontology and dynamics have to be linked with an appropriate probability measure. Thus, the question is: What is the law that describes the individual processes that occur in nature (dynamics) and what are the entities that make up these individual processes (ontology)? \citep[222]{esfeld2019measurement}
\end{quote}
In the same vein, Maudlin writes:
\begin{quote}
    A physical theory should contain a physical \textit{ontology}: What the theory postulates to exist as physically real. And it should also contain \textit{dynamics}: laws (either deterministic or probabilistic) describing how these physically real entities behave. In a precise physical theory, both the ontology and the dynamics are represented in sharp mathematical terms. But it is exactly in this sense that the quantum-mechanical prediction-making recipe is not a physical theory. \citep[p.~4, original emphasis]{maudlin2019quantumphil}
\end{quote}

So let's go after this ontology. Because we stated the measurement problem as concerning an articulation of the wave function,\footnote{~That is not the only possible way to formulate quantum mechanics; for a critical view on the subject, see \citet{bokulich2020wavefunction}; for formulations other than standard quantum mechanics, see \citet{styer2002nineqm}. For the sake of simplicity, we focus on the ontological articulations of wave functions, but our point could be equally framed for other formulations too.} we can start from there: assuming the wave function as a basic entity, common to the three subsequent theories. So, besides the wave function, here's an ontological showcase.

\begin{itemize}
    \item The Bohmian QM$_{hid}$ postulates \textit{point particles} as fundamental ontological entities of QM \citep[see][220]{bohm2006undivided}. Such point particles always have determinate positions that move according to the non-local pilot wave, albeit we are unable to know them beforehand since they are ruled by hidden variables.
    
    \item The GRW versions of QM$_{col}$ can be seen both as a theory of the wave function and as a particle ontology, dubbed as ``GRWp'', with particles moving on continuous trajectories \citep[see][]{allori2021grw}. It can be, however, also associated with a continuous matter density ontology, in the literature called ``GRWm'', in which its ontology consists of localized entities in a three-dimensional physical space \citep[see][]{GRW, durrlazarovici2020understandingqm}; and also as an ontology of a collection of instantaneous ``flashes'', dubbed ``GRWf'' \citep[see][]{bell2004speakable}.\footnote{~For details, see \citet[156]{esfeld2018naturalized}.}
    
    \item The many-worlds interpretation of QM$_{bra}$, or Everettian Quantum Mechanics (EQM), can be seen as a description of physical reality in terms of the wave function alone \citep[see][48]{durrlazarovici2020understandingqm}. But also with an ontological addition in which splitting universes are posited \citep[see][]{dewitt1971many}, or without such a reification of these universes, understood as emergent entities \citep[see][]{wallace2012emergent}.
\end{itemize}

So we have to interpret quantum mechanics to see what it tells us about what there is \citep[see][152]{esfeld2018naturalized}, and to ``interpret'' it is to come up with a solution to the measurement problem. All of these quantum theories sketched above are usually considered quantum theories that can be interpreted realistically since they solve the measurement problem. Even in the literature that does not admit the measurement problem, mentioned above as the ``problem of outcomes'', this problem is understood as the ``problem of superposition'' \citep[see][145]{deronde2019probing}, which can be stated most simply with the following question: \textit{why do we never experience a superposition}? In Schrödinger's famous cat experiment, where the quantum-mechanical description of the cat, before measurement, is given in terms of a superposition between the ``awake'' and ``asleep'' state vectors, the explanation for each solution is as follows:

\begin{quote}
    In fact in the pilot-wave theory [QM$_{hid}$], we do not observe macroscopic superpositions because the complete description of the system is given by the specification of particles and wavefunction, and particles are always localized, just like cats, whether they are [asleep] or [awake]. In the many-world theory [QM$_{bra}$] the various terms of the superpositions `live' in other worlds which are inaccessible to us and do not interact with us, and this explains why we do not encounter the [awake] counterpart of a [asleep] cat. Finally, in the spontaneous localization theory [QM$_{col}$] the fact that the wavefunction localizes very fast for macroscopic objects explains why we never see macroscopic objects in superposition states, and why [asleep] cats remain, perhaps unfortunately, [asleep]. \citep[75]{allori2021grw}
\end{quote}

Usually, the focus is on underdetermination: we have \textit{at least} three distinct theories that deal with quantum phenomena, all of them equally empirically successful, and with different ontological outputs. Currently, we have no way to solve the problem of theoretical (and consequently ontological) underdetermination \citep[see][pp.~231--232]{durrlazarovici2020understandingqm}.

\subsection{Naturalized Ontology}

To many authors, contemporary analytic metaphysics is worthless as a guide to the objective features of the world \textit{unless} it can be continuously guided by contemporary science \citep{Ladyman2007evmustgo}. Such a project is known as the \textit{radical naturalization} of metaphysics, which advances a skeptical attitude towards most of what currently falls under the label of ``analytic metaphysics''. This kind of debate suffers from some ambiguities in its scope and aims, a difficulty which could be at least partly overcome (we believe) if a proper distinction between ``ontology'' and ``metaphysics'' is advanced. Such a careful distinction could avoid many problems we find in the current literature on the naturalization project.

One of the most common of such problems is the naturalistic expectation that scientific theories provide everything we need to know in terms of metaphysics and ontology, so that we can \textit{read off}, \textit{extract}, or \textit{derive} a metaphysics (again, a term often treated as interchangeable with ``ontology'') directly from a given theory's scientific basis. If for a moment we dissociate the two terms, we might be able to see the achievements and limits of the radical naturalization project (and this could work to the benefit of such a project). Here's Hofweber:

\begin{quote}
In metaphysics we want to find out what reality is like in a general way. One part of this will be to find out what the things or the stuff are that are part of reality. Another part of metaphysics will be to find out what these things, or this stuff, are like in general ways. Ontology, on this quite standard approach to metaphysics, is the first part of this project, i.e. it is the part of metaphysics that tries to find out what things make up reality. Other parts of metaphysics build on ontology and go beyond it, but ontology is central to it, \textelp{}. Ontology is generally carried out by asking questions about what there is or what exists. \citep[13]{Hofweber2016MetOnt}
\end{quote}

Given that distinction between ontology and metaphysics,\footnote{~Similar distinctions can be easily found in the literature. See, for instance, \citet{arenhart2012ontological, fine2017, thomsonjones2017existencenature, arroyo-arenhart2022foop}.} it seems that naturalists may have reason to appreciate ontology as properly continuous with science, while the same is not so evident with metaphysics. After all, as we saw, we can extract ontological commitments from the theories. One may argue that we can extract more than one ontology from the same family of theories, but our point here is not underdetermination, but the \textit{ontological extraction}. Consider this claim:

\begin{quote}
    Metaphysics is ontology. Ontology is the most generic study of what exists. Evidence for what exists, at least in the physical world, is provided solely by empirical research. Hence the proper object of most metaphysics is the careful analysis of our best scientific theories (and especially of fundamental physical theories) with the goal of determining what they imply about the constitution of the physical world. \citep[104]{maudlin2007metaphysics}
\end{quote}

The above quote essentially distills what we call the extractivist conception of ontology, according to which we can extract the ontology of science. We should mention, in passing, that this is by no means the only strategy; on a diametrically opposite side, we have the conception called ``Primitive Ontology'', whose methodology is reversed: instead of obtaining the ontology as derived from a scientific theory given beforehand, it starts with a previously assumed ontology and attempts to build a physical theory that correctly describes it: ``\textelp{} when a scientist proposes a fundamental physical theory, she already has in mind what the theory is fundamentally about: the primitive ontology''
\citep[61]{allori2013primitive}.

We will not deal with this proposal here, as we are interested in the debate with naturalists, who try to inherit some epistemic credentials from science to ontology, with ontology \textit{extracted} from science. It does not seem to make sense to put an ontology beforehand if what is being sought is to attribute some epistemic respectability for ontology as \textit{derived} from science. So, in the sense of obtaining an ontology adopted by the extractivists, is the task of the \textit{ontologist} over when an interpretation of quantum theory is adopted? According to the meaning of the term 'ontology' discussed earlier, we suggest that it is possible to extract the ontologies from the different interpretations of quantum mechanics, that is, we extract from the scientific theories a description of what exists according to the scientific theory in question. But then again: is it really possible? And: is it all we have to do?

In order to understand the limits of naturalization, it is useful to introduce here a distinction between ``shallow'' and ``deep'' versions of scientific realism \citep[see][]{French2018RealMetaph}. The shallow-type scientific realist is the one that rests content with repeating the very same declarations of scientific theories when it comes to discussing ontology; she would be comfortable sitting near the ontology campfire because \textit{at least ontology} is fully granted by science, thus inheriting its epistemic credentials. But is that all? One could go further and ask additional questions, willing to understand what kinds of entities are those that are being posited by our best scientific theories. In ontological terms, what are such posits? Are they objects with properties (substances), structures, dispositions, facts, or perhaps even something else? As we argue, this remains at the ontological level---hence, somewhat shallow to deep realists\footnote{~Deep realists believe that one cannot be a realist unless metaphysical questions---\textit{qua metaphysics}---are properly answered. That is, one must properly spell in detail the kind of entities scientific theories are committed to, e.g. if they're objects, are they individuals or not? If they're structures, what is their metaphysical profile: universals or tropes? And so on. As there is no debate on whether such deep metaphysical questions are to be settled within scientific theories \citep[see][chap.~4]{KraFre2006}, we focus solely on the shallow spectrum of scientific realism \citep[see also][]{arenhartarroyo2021veritas}.}---but we also argue that the kind of ontological answers that these questions pose is not to be found within scientific theories themselves. Let us dwell on that for a moment.

\section{Two senses of ontology}\label{paper:ont:sec-4}

We can list, à la Quine, what exists according to a theory. Suppose we are talking about quantum mechanics, and find that the theory requires the existence of electrons: we then extract commitment with an entity from quantum mechanics. Has ontology been naturalized? So far, it seems so. But is this all the work of ontology, i.e. the establishment of the catalog of each interpretation of quantum mechanics and leaving it as it is?

Our claim is that this is not everything if one is to go beyond shallow realism. The literature is elusive in this respect, giving us only hints about what else is typically done by the scientific ontologist. For instance, we find in \citet[p.~30, emphasis added]{ney2014metaphysics} that ``ontology'' is ``1. the study of what there is''; but also ``2. a particular theory about the \textit{types} of entities there are''. In this sense, there would be two complementary tasks of ontology: at first, the establishment of an ontological \textit{catalog}, which says what things exist (tables, chairs, electrons, Everettian worlds, Bohmian pilot waves, etc.); and secondly, the establishment of ontological categories, according to which we can classify what \textit{type} of entities they are (objects, properties, relations, events, processes, immanent powers, structures, wave functions\dots) and, eventually, investigate into whether some type is more fundamental than other(s). Quine himself, while discussing ontology, sometimes invites confusion between the two types of questions: 
\begin{quote}
\textelp{} we are quite capable of saying in so many words that \textit{there are} black swans, that \textit{there is} a mountain more than 9000 meters high, and that \textit{there are} prime numbers above a hundred. Saying these things, we also say by implication that there are physical objects and abstract entities; for all the black swans are physical objects and all the prime numbers above a hundred are abstract entities.
\citep[67]{quine1951commitment}
\end{quote}

Here, besides going from entities of the catalog to their types, the plan is that one can also \textit{infer} on their specific place in the concrete/abstract divide. The latter is a discussion we shall not enter into here. Now, on what concerns the possibility that swans are objects and that this is inferred from their existence, the additional premise ``all black swans are physical objects'' must certainly be added, and this is perhaps what is missing most of the time (as we shall point to). Indeed, this additional step in the inference connecting an entity to its type is precisely what is not typically provided by quantum mechanics, in the case of the catalog it provides. Having the catalog is not, \textit{eo ipso}, having the type of each entity in the catalog. Using a similar vocabulary, \citet[49]{bueno2023apq} has also claimed that ``facts, substances and structures'' are ``ontological categories''. Now, the fact that Bueno himself is eliminativist towards the reification of such categories \citep{bueno2023apq,bueno-etal2015} matters little to our goals; what is at stake here is the distinction between the two parts of ontology we have highlighted and their methodological relation with science.

For terminological purposes, whenever a fixed discourse or theory is assumed as given in the background, we shall call the catalog of entities it postulates by ``$\mathscr{O}_{cat}$'', and we shall call  by ``$\mathscr{O}_{typ}$'' one answer to the question concerning the type of entities this catalog is classified into. In this distinction, only the first part of the ontological project ($\mathscr{O}_{cat}$) could be, in principle, directly naturalized. So establishing the catalog for each interpretation is a relatively passive job of the scientific ontologist: just note which entities the theory says exist. Easy peasy. Now comes the creative part ($\mathscr{O}_{typ}$): classifying these entities into ontological categories. Note that this creativity moves science away from ontology and is no longer a naturalizable aspect. This is because the theory itself does not tell us the type of entities with which it is committed ontologically. Let's take a closer look at that.

It is quite standard to state that if we have an object-oriented ontology, metaphysical underdetermination concerning individuality arises: if quantum entities are \textit{objects}, they can be understood within a metaphysical framework of individuals or non-individuals \citep[see][chap.~4]{KraFre2006}. The metaphysical underdetermination concerning individuals and non-individuals is one of the reasons why structural realists think that we should shift our ontological basis, from an object-oriented ontology to an ontological basis in which such metaphysical underdetermination does not occur \citep[pp.~419--420]{ladyman1998StrucReal}. The proposed change of ontological basis would recommend a commitment to a structure-based ontology. This position is known in the literature as ``ontic structural realism'' (OSR). Understood as an eliminative proposal about objects, the view suggests that entities ontologically considered ``objects'' carrying certain properties would simply not exist, but only structures understood as collections or families of relations. OSR then takes the notion of structure ``\textelp{} to be primitive and ontologically subsistent'' \citep[420]{ladyman1998StrucReal}. Hence, in the ontological ($\mathscr{O}_{typ}$) terms, eliminative OSR would claim that ``\textelp{} relations exist, and can be ontologically fundamental, without there being relata'', and ``\textelp{} objects become otiose, and are eliminated from the basic ontology'' \citep[262]{friggvotsis2011structural}. In this sense, the items in the catalog that were obtained by the naturalized part of the ontology ($\mathscr{O}_{cat}$)---e.g. electrons---would be interpreted as being items of the type ``structures'' in $\mathscr{O}_{typ}$. And in doing so, it is claimed that issues concerning individuality no longer appear---this would be a metaphysical issue exclusive for object-based $\mathscr{O}_{typ}$.

Another way of framing the distinction between the two tasks of ontology concerns understanding the reference of ``electrons'' in terms of the reference of mass terms, and not in terms of objects (which are themselves the reference of count terms) or of structures. Rather than being things (as per an object-oriented ontology) or nodes of relations (as per a structure-oriented ontology), electrons would be something more like stuff, that comes in quantities, not in units. In that sense, they are much more like water quantities than unitary particles \citep[for the details of this proposal, which we omit for the sake of space, see][]{Jantzen2019}. What is relevant here is that electrons, as items in the catalog of the ontologist, are understood as a distinct type when it comes to $\mathscr{O}_{typ}$. That is, for one answer to $\mathscr{O}_{cat}$ there corresponds at least three distinct possible answers to the questions concerning $\mathscr{O}_{typ}$.  

If we stick with the distinction between ontology and metaphysics, we will divide the work of ontologists and metaphysicians, respectively, as this two-part job: 1) tell what entities exist; 2) tell what those entities are. Part 2) would be aptly called the non-naturalizable task of metaphysics, or the ``deep'' part of scientific realism, to use the terminology offered by \citet{magnus2012} and \citet{French2018RealMetaph}. But this is not our subject; we are dealing exclusively with ontology. There seems to be a more fundamental distinction in the ontologists' part of the job. Notice: an electron is an entity, but we can classify it within several types, without yet getting into metaphysics.

Let us concentrate on electrons again to put that point in clear terms. Consider the following claim: ``quantum mechanics is committed with the existence of electrons''. This answers the ontological question related to the \textit{catalog} of the theory. After that, there is the problem of the \textit{type} of entity it is. For example, we can say that they are objects. Alternatively, one can say that they are structures: if we were ontological structuralists, we would recognize that basic entities would be structures, not objects \citep[see in particular the articulation of the view in][]{Saunders2003-SAUSRA}. Structures and objects are still in the ontology. But it seems they are in a different part of the ontology than the ontological inventory or catalog, which lists the entities existing inside the theory. In this way, quantum mechanics exhibits this further ontological underdetermination. Not only of the catalog---given that different interpretations furnish different answers to $\mathscr{O}_{cat}$, as we have seen---, but also the ontological underdetermination of the catalog's typification: electrons, for example, can be considered structures or objects, and this typification is not extracted from the theory.

In the next step, one could \textit{metaphysically} understand these objects as individuals or as non-individuals---this being the traditional way of spelling out the notion of ``metaphysical underdetermination'', which emerged in this context \citep[again, see][]{KraFre2006}. Of course, we should mention again that the theory isn't also informative about the catalog's metaphysics; viz., it doesn't tell us whether electrons \textit{qua objects} are individuals or non-individuals. A similar underdetermination appears for other kinds of answers to the $\mathscr{O}_{typ}$ question; for instance, if one considers electrons as structures, quantum mechanics is silent about the metaphysical profile of electrons \textit{qua} structures \citep[see][]{arenhartbueno2015}, so there is this recurring gap. But recall that this additional metaphysical layer, concerning the question of nature, is not our goal here. Rather, we're pursuing a methodological distinction between two kinds of ontology.

Now let us reintroduce the solutions for the measurement problem in this discussion. Apart from ``electrons'', each of them posits different stuff in their respective $\mathscr{O}_{cat}$. Everettians, for example, are known to populate the ontology with many worlds; the many-worlds interpretation of Everettian quantum mechanics forces that upon them. Once this solution to the measurement problem is assumed, there's no way around such an ontological ($\mathscr{O}_{cat}$) posit. Recall that one way of solving the measurement problem is by appealing to the existence of such worlds, so the existence of Everettian worlds is an ontological consequence of such a solution. Now whether one wishes to interpret Everettian worlds as OSR ``structures'', i.e., as relations without relata \citep{french2014structure}; or as ``objects'', i.e. entities endowed with an individuality profile \citep{brading-skiles-2012} is a matter about which Everettian quantum mechanics is silent. In this way, some can support both EQM and structural realism, just as \citet{wallace2012emergent} does. Or EQM and entity realism.\footnote{~And just like in the previous case---viz., of standard QM and electrons---EQM will be also silent on metaphysical matters, e.g., whether Everettian worlds are concrete worlds à la Lewis or abstract worlds à la Plantinga\citep[for a discussion on these matters, see][]{arroyo-arenhart2022foop}.} This is so because the naturalization goes only up to $\mathscr{O}_{cat}$; $\mathscr{O}_{typ}$ is not determined by EQM.

The same thing can be said about Bohmian mechanics, and here things are even more troublesome.\footnote{~We express our appreciation to an anonymous reviewer for pressing us on this point.} For it has been claimed that a solution to the measurement problem yields a \textit{structured} ontology, i.e., a combination between   $\mathscr{O}_{cat}$ and $\mathscr{O}_{typ}$ \citep{lopez2023}; for instance, Bohmian mechanics is committed not only to particles and pilot waves in the $\mathscr{O}_{cat}$. It is \textit{also} committed with the particles being objects in $\mathscr{O}_{typ}$---moreover, \textit{individual} objects, hence determining not only both parts of ontology but also its metaphysical aspect. If that were the case, ontology and metaphysics would be completely determined by Bohmian mechanics, hence fully naturalizable.

We resist such a claim by pointing out to the fact that, notably, Bohmian mechanics can also be interpreted within OSR \citep[see][]{lam2015, Esfeld2017-ESFHTA}. More specifically, as \citet{lorenzetti2022} put it, in the extant literature there are three ways\footnote{~In addition to a fourth way, which is the coherency-based OSR approach \citet{lorenzetti2022} himself develops in the aforementioned article.} of interpreting QM$_{hid}$ in structural terms:
\begin{quote}
(i) according to the eliminativist position, the wave function is all there is; (ii) priority-based OSR claims that the Bohmian particles are determined by the wave function; and finally, (iii) according to moderate OSR the wave function and the particles are both fundamental and they are inter-dependent. \citep[12]{lorenzetti2022}
\end{quote}
The first option is eliminativist not just with regards to particles, but also with \textit{objects} altogether, hence it would be even more precise to state that ``according to the eliminativist position, the wave function [qua structure] is all there is''. The priority-based and moderate OSR admits objects and structures within their $\mathscr{O}_{typ}$. Once again, a fixed $\mathscr{O}_{cat}$ (e.g., Bohmian particles and pilot waves) can be interpreted with different $\mathscr{O}_{typ}$, viz., as structures, objects, and so on.

Different $\mathscr{O}_{typ}$ ontologies are not being ``read off'' from each solution to the measurement problem in the same sense of $\mathscr{O}_{cat}$. In the latter case, ontological extraction seemed easier: a basic catalog was `extracted'/`derived from'/`read off' for each quantum interpretation as a consequence of the direct postulation of such entities for a solution to the measurement problem. But with $\mathscr{O}_{typ}$ the situation is different: each of the proposals $\mathscr{O}_{typ}$ listed above should now be seen as being plugged in on the top of a given catalog, such catalogs being compatible with distinct ontological types. The link Quine saw between swans and objects is not delivered by our favorite theory: one may connect them with objects, structures, processes, and so on.

To emphasize that point, here's Quine again:

\begin{quote}
    For let us reflect that a theory might accommodate all rabbit data and yet admit as values of its variables no rabbits or other bodies but only qualities, times, and places. The adherents of that theory, or immaterialists, would have a sentence which, as a whole, had the same stimulus meaning as our sentence ``There is a rabbit in the yard''; yet in the quantificational sense of the words they would have to deny that there is a rabbit in the yard or anywhere else. \textelp{} prima facie, are two senses of existence of rabbits, a common sense and a philosophical sense. \citep[98]{quine1969exist}
\end{quote}

The point we wish to stress in this passage is the independence between the catalog and its type. Quine is again mixing catalogs and types. While the verbal input (the evidence available) suggests that there is a rabbit, making for an agreement regarding the range of quantification, there is also another way of cutting what is going on according to which there is disagreement, which is related to the type. Without making it completely explicit, Quine reads `rabbits' as pertaining to objects when it comes to the type of ontology. However---as he himself also recognizes---that evidence is compatible with other options for types, related to properties instantiated in space and time. Our point is that this latter part (either object, or else properties) does not come with the rabbit out of the evidential hat. By quantifying over rabbits, we are not quantifying over objects, or properties. The decision concerning the types comes only at a later stage. In other words, we are claiming that when doing Quinean-like ontology, one is not quantifying over the ontological categories($\mathscr{O}_{cat}$), but over the ontological catalog. Ontologists wouldn't---shouldn't!---list `objects' or `structures' among their ontological catalog, as if there were $\{\text{persons, tables, chairs, electrons, objects,}\dots\}$. Lowe puts the issue in the following terms:

\begin{quote}
    I should perhaps remark, indeed, that ontological categories are not themselves to be thought of as \textit{entities} at all, nor, \textit{a fortiori}, as comprising a distinctive ontological category of their own, the category of \textit{category}. To insist, as I do, that ontological categories are categories of being, not categories of thought, is not to imply that these categories are themselves \textit{beings}. \citep[p.~6--7, original emphasis]{lowe2005}
\end{quote}

\begin{figure}[ht]
\centering
\begin{forest} for tree={grow=-90}
[Measurement problem
[QM$_{hid}$
[Bohmian particles 
[Objects ]
[Structures ]
] 
]
[QM$_{col}$
[GRW particles
[Objects ]
[Structures ]
]
]
[QM$_{bra}$
[Everettian worlds
[Objects ]
[Structures ]
] 
]
]
\end{forest}
\caption{The measurement problem, quantum interpretations, $\mathscr{O}_{cat}$, and $\mathscr{O}_{typ}$.}
\label{fig:ontologicalmap}
\end{figure}
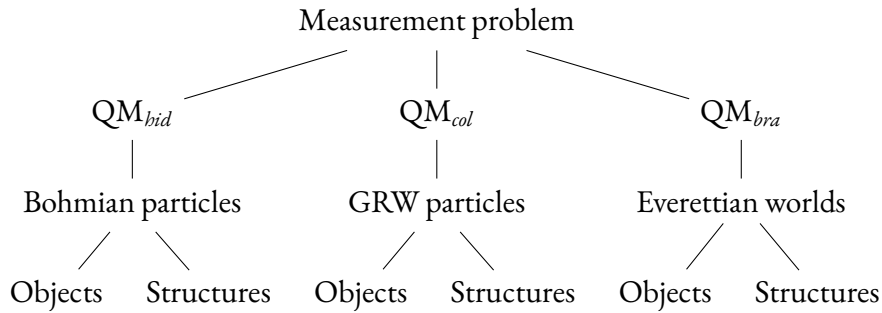

Our way out is to stress that a---naturalistically-approved---\textit{derivation} is available only to $\mathscr{O}_{cat}$. It is not possible to find the accompanying middle premise in analog lines to ``all black swans are physical objects'' in quantum mechanics; that is, quantum theory does not tell us that ``all electrons are physical objects'', instead of  ``all electrons are nodes in a structure''. With that distinction in mind, the term ``object'' would no longer be understood as ontologically neutral. For that, we would need to choose one, as ``entity''. This is because the term ``object'' already buys the package of an object-based ontology (to be contrasted with a structure-based ontology, for example). This type of care has gained attention; we see this, for example, in \citet[2]{meincke2020dispositionalism}: ``I deliberately speak of ``entities'' rather than (as many scholars do) of ``objects'' to leave open the possibility that also non-object-like entities, such as processes, events, states of affairs, structures etc.''. Then we would need to distinguish between the nature and the classification of the entities' modes of existence. A graphical resource (Figure \ref{fig:ontologicalmap}) might help us to visualize the methodological distinction we've been proposing thus far.

We should mention that while $\mathscr{O}_{cat}$ was kept fixed in the above example, this might not be the case. The ontological content, qua $\mathscr{O}_{cat}$, of each quantum interpretation, is not clear. For instance, we've seen that it is not settled whether GRW should be conceived as a theory of particles, mass densities, or flashes; the same could be said about Bohmian mechanics and Everettian quantum mechanics: respectively, whether there are just one or many particles; whether there are many worlds, and so on. It should also be noted that the list of $\mathscr{O}_{typ}$ is not exhausted by objects and structures; one may add processes, events, and so on.\footnote{~Figure \ref{fig:ontologicalmap} is very limited in this sense. As a graphic resource, however, we deliberately chose to enhance clarity over generality.} That said, it should be clear now that one might be an Everettian who proudly admits that ``there are multiverses according to EQM''. And that the dispute on the existence of many worlds in EQM is, in our terms, an ontological dispute in the sense $\mathscr{O}_{cat}$. A further dispute among those who admit the existence of multiverses in EQM is whether we should understand these universes as objects, as structures, or as something else. In our terminology, the dispute now is about ontology in the sense comprised by $\mathscr{O}_{typ}$.

Moving ahead, the metaphysical question qua metaphysics, viz. about the metaphysical \textit{nature} of such worlds is subsequent to $\mathscr{O}_{cat}$ and independent of $\mathscr{O}_{typ}$ in a logical sense. Take, for instance, the quandary between Everettian actualism \textit{versus} quantum modal realism \citep[see, for example][]{Conroy2018actualism, wilson2020QMmodal}, the former defending that Everettian worlds should be metaphysically understood in actualist terms, viz., as non-existent entities, and the latter defending a modal-realist account of Everettian worlds, viz., as existent entities \citep[for this discussion, see][]{arroyo-arenhart2022foop}. These are questions regarding the very ontological existence of Everettian worlds that should or shouldn't be admitted in one's $\mathscr{O}_{cat}$ ontology; but this debate seems to be largely independent on where one places them within $\mathscr{O}_{typ}$ ontology.\footnote{~A similar discussion, illustrating the use of the conceptual distinction introduced by our framework appears in \citet{Esfeld2017-ESFHTA}, where structural realism (a type ontology) is put to work together with an interpretation of quantum mechanics by primitive ontology. In a sense, it reinforces the point that ontology as providing a catalog through an interpretation (the primitive ontology) is a different business than providing for the type of entities posited.}

Bluntly put: one might be (say) a quantum-modal-structural-realist or a quantum-modal-\textit{objectual}-realist about Everettian worlds, but there seems to be no point in discussing quantum modal realism once Everettian worlds are not even admitted in $\mathscr{O}_{cat}$---e.g., cases in which one adopts pilot-wave or collapse-based quantum theories.

As a consequence of the independence of metaphysics qua metaphysics and ontology qua $\mathscr{O}_{typ}$, one may ask about the metaphysical nature of what is admitted in $\mathscr{O}_{typ}$, just as it was the case discussed above concerning the metaphysical nature of the entities admitted in $\mathscr{O}_{cat}$. It thus remains to be specified what an object is, or what a structure is, in metaphysical terms. 

Saying, for instance, that ``a structure is not an object'' does not help much. In fact, this is a problem of structural realism: it does not have a ``metaphysical profile'' for structures---we don't know what they are, so how can we be a realist about them in a ``deep'' sense \citep[see][]{arenhartbueno2015}? But of course, there may be one, and that is a question for clever metaphysicians. But to say whether what exists, exists according to the ontological category of objects or structures, is a question for clever ontologists (and these two may overlap, of course---they're clever \textit{philosophers} after all). For those willing to go into ``deep'' metaphysical questions: what is the nature of these entities that populate the catalog of scientific theories? From what has been discussed so far, we can even learn about the field of possibilities of metaphysics when we take ontology and science into account \citep[see][]{RaoJonas2019dualismQM}; however, just as physics does not determine metaphysics \citep[see][]{KraFre2006}, the lesson about ontology is also clear: ontology (whether $\mathscr{O}_{cat}$ or $\mathscr{O}_{typ}$) does not determine metaphysics.

\subsection{Objections}
Below we list worries one might have by reading our proposal, and how we address them.

\paragraph{1. The problem of underdetermination.}
 Given that underdetermination is a problem, how can one learn ontological lessons (\textit{qua} $\mathscr{O}_{cat}$) from QM?\footnote{~Indeed, two anonymous referees raised such an objection. We thank them for prompting a closer examination of this matter.}

To answer questions of this kind, one must disentangle naturalism from scientific realism. It is true that one currently has no intra-theoretical grounds to choose between QM$_{hid}$, QM$_{col}$, and QM$_{bra}$. It is \textit{also} true that one has more than one ontological output (again, \textit{qua} $\mathscr{O}_{cat}$) from each one of these QM. To stick with just one example: within QM$_{bra}$ one has no grounds to decide between two different $\mathscr{O}_{cat}$, say, the catalog of many \textit{minds} and the catalog of many \textit{worlds} \citep{dewitt1971many,lockwood1989}. Even within the many worlds $\mathscr{O}_{cat}$, there's still no consensus on whether they should count as concrete or abstract \citep{arroyo-arenhart2022foop}. In section \ref{paper:ont:sec-3} we've mentioned something similar to QM$_{hid}$ and QM$_{col}$.

From the methodological point of view that we stressed here, however, underdetermination should not scare naturalists. Methodologically, naturalists should care solely about the fact that $\mathscr{O}_{cat}$, whatever it may come out to be, was \textit{provided} by science. It is an output from science, not an input from a philosopher. The way we understand the naturalist project---viz., the project of having philosophical content derived from science---that's what counts: being a derivation, not an addition. The fact that science does not determine yet all facts concerning ontology is not a problem for a naturalist; the problem is only when philosophers, qua philosophers, opt for deciding for science. 

Let us press this a little further. From a methodological point of view, $\mathscr{O}_{typ}$ questions do not receive the same kind of support as $\mathscr{O}_{cat}$ questions. Ontological typification is largely independent of ontological cataloging. $\mathscr{O}_{cat}$ is derived from science, and, in the sense we understand `naturalism' (viz. `derived from' science) is the only naturalizable part.

Here we can draw a parallel between $\mathscr{O}_{typ}$ ontological questions and distinctively metaphysical questions, e.g., questions about the individuality profile of quantum objects. As \citet[21]{esfeld2013} puts it, ``\textelp{} attributing a primitive thisness to objects is a purely metaphysical move that one can always make, physics be as it may''. We might add, \textit{not} attributing a primitive thisness to objects, i.e., sustaining a metaphysical profile of non-individuals, is as well \citep[see also the discussion in][]{arenhartarroyo2023roads}. The methodological point at issue is this: whatever the metaphysics, there is no difference in physics. This is also the case with $\mathscr{O}_{typ}$ categorizations. Whatever the categorization, e.g. $\mathscr{O}_{typ}$ of structures or $\mathscr{O}_{typ}$ of objects, the quantum interpretations remain unchanged.

An interesting question is whether there is an empirical difference if there is a difference in the catalog? For example, in a $\mathscr{O}_{cat}$ ontology of many worlds, probability needs to be explained; in a $\mathscr{O}_{cat}$ ontology of GRW flashes, the new constants of nature need to be explained; if we have a particle guided by a pilot wave, locality---and consequently the connection with relativistic theories---needs to be explained. Whatever the answer to this question, the situations we have listed are scientific problems, and we can expect the ontological catalog to be shaped by how science addresses them. From a methodological point of view, however, the same cannot be expected for ontological questions of the $\mathscr{O}_{typ}$ type (and not even distinctly metaphysical questions).

In any case, such questions are independent of realist or antirealist attitudes towards their underdetermination. As it is well-known, underdetermination may be a feature or a bug, depending on how one stands between the epistemic stances of realism and antirealism. And the matters of naturalization do not bear on them: if science points to an actual plurality of accounts, it is not the job of the naturalist to choose `the correct one' for science. According to \citet[1850]{guaypradeu2020}, it is an ``oversimplification'' of naturalism ``to reduce science-based metaphysical approaches to realism''. Of course, they're not considering the distinction between metaphysics qua ontology and metaphysics qua metaphysics that we're assuming here; neither they are considering the further distinction between metaphysics qua ontology qua catalog (sic) and metaphysics qua ontology qua type (sic!). The moral of the story, however, is similar: that the outcomes of such a metametaphysical (qua metaontological) discussion shouldn't depend on one's epistemic stance towards scientific realism or antirealism.\footnote{~In broad strokes, this aligns with what \citet[chap.~2]{emery2023} recently argued in length, viz., that nothing in naturalism requires the specific alignment towards scientific realism--antirealism. Be it content/ontological naturalism or methodological naturalism; be it radical, moderate, or limited naturalism; virtually nothing in the characterization of naturalism depends on a specific epistemic stance on scientific realism--antirealism matters (excluding, of course, the antirealist qua instrumentalist stance).}

\paragraph{2. The Quinean principle is not the only game in town; why use it without mentioning others?} Indeed, at least three other metaontological principles could be used in discussing quantum ontology.

\begin{itemize}
    \item The Easy Ontology approach \citep[6]{Thomasson2009-THOTEA-3}: ``on this view existence claims (whether formulated in English or quantificationally) are only truth-evaluable when paired with a term or terms that come with application conditions, so that their truth may be evaluated by way of establishing whether or not those application conditions are fulfilled.''
    \item The dynamical principle \citep[186]{north2013}: ``dynamical laws are a guide to the fundamental nature of a world''.
    \item The grounded ontological commitment \citep[91]{imaguire2018}: ``only sentences that express fundamental facts, i.e. facts that do not have additional grounds, determine the furniture of reality via ontological commitment''; one should ontologically commit oneself only to the fundamental entities.
    \item The minimal divergence norm \citep[564]{emery2017}: the ``fit between the way a theory says the world is and the way the world appears to be''; our immediate intuitions should guide quantum ontology, and they should diverge as minimum as possible.
\end{itemize}

We didn't use any of them in developing our distinction between $\mathscr{O}_{cat}$ and $\mathscr{O}_{typ}$. Here's why, beginning with the latter.

As \citet[p.~564, emphasis added]{emery2017} herself acknowledges, the simplicity of her metaontological principle is a principle for ``scientific \textit{theory choice}''. It is a principle that could be called upon to choose between different ontologies (qua $\mathscr{O}_{cat}$) to fill the scientific realists' goal. But it is not a metaontological principle for obtaining a $\mathscr{O}_{cat}$ from a scientific theory. The Quinean principle of ontological commitment is precisely that. While an important theme, theory choice in quantum ontology is beyond the scope of the current paper, which is focused solely on the methodology of obtaining (not choosing) an ontology from science.

North's and Imaguire's proposals, however, are a different matter. They announce their metaontological principles as an ``updated version'' \citep[p.~189, fn.~10]{north2013} or a modification \citep{imaguire2018} version of the Quinean ontological commitment, largely used in this paper. Recall that we distinguished between ontology as catalog and ontology as type investigation. Opting for the above-mentioned metaontological methods for obtaining an ontology would generate some tensions and unanswered questions concerning the relation between ontology understood as catalog and ontology understood as type. Fundamental entities may be understood as a type or as something present in the catalog. If fundamental entities come from the catalog, how did we manage to obtain the catalog to begin with? It seems that the approach under consideration would have to be elaborated over an already existing account to generate the catalog, so that we would be able to separate afterwards what is fundamental and what is not. If the fundamentals come from the type ontology, then, again, we depend on a previous catalog, which somehow instructs us as to the type of entities that are required as fundamentals (recall the discussion between object-oriented ontologies and structure-oriented ontologies: given a catalog comprised of protons, electrons and the such, we need to sort out the type of such entities). 

An additional difficulty is that determining what is fundamental, in any of the senses, is very far from being an issue that can be solved by quantum mechanics itself. So, although one may elaborate on such an approach, it is still very difficult to see how it can appeal to naturalism in ontology. This is why we opted for the unmodified, old-fashioned Quinean metaontological method. More specifically: a) the method proposed by \citet{imaguire2018} would generate commitment with a previous approach to ontological extraction and a high dose of ontology-first approach to ontology---hence, unwanted for naturalism right from the start; b) the method proposed by \citet{north2013} was recently shown to be able to support several quantum ontologies \citep{matarese2022}, so, in a sense, it is still a bit underdetermined when it comes to determining exactly claims about $\mathscr{O}_{cat}$. Our claim, however, is more general than that, viz., on the prospects of naturalizing ontology as a whole, category and types included. 

Thomasson's approach, on the other hand, is a sort of flexibilization of the Quinean approach, and would deliver very similar results, by a kind of different method. The major differences lie in her not being committed to some of the traditional Quinean tenets, like Quine's extensionalism.

All in all, Quinean methodology seems to be part of the orthodoxy when it comes to naturalism. Certainly, things can change, and our discussion and distinctions between categories and types would carry over to any other method naturalists prefer to extract their catalogs. Nothing of importance in our discussion depended on our choice of the Quinean methodology, although we acknowledge that it seems the most suited for naturalists so far.

\section{Conclusion}\label{paper:ont:sec-5}

The naturalization project is undeniably attractive: ``what if we could transfer the epistemic respectability of science to philosophical disciplines?'', naturalists wonder. However, things are not that simple. Some disciplines resist naturalization and claim creative independence. As we have argued, this is the case with a great part of the problems involved in ontology. 

When philosophers say that Quine alone ``saved metaphysics'', the discipline in question that was ``saved''---in the biased sense of having been made respectable for maintaining close epistemic relationships with science---was ontology. In traditional Quinean metaontology, this means: ``what is there, according to such and such scientific theories?'' and, as the one who informs us about this catalog is science itself, many thought that the project had been completed, and that ontology would have been successfully naturalized.

Perhaps that part has in fact been successfully naturalized. After all, we can literally read the ontological commitments of scientific theories. The problem of having many theories competing with each other for a (roughly) true description of the world turns out not to be a problem for naturalism itself, but a problem for scientific realism (underdetermination), but the point of naturalists, which we call ``extractivism'' is untouched: all that matters is that ontology comes from science; if we have a lot of ontologies because we have a lot of scientific theories, well, that's kind of a problem for science (and not for ontology). If there is no single best scientific theory from which one can extract one's ontology, at least the epistemic problem of ontology is properly addressed.

However, when we realize that the work of ontologists does not end with the establishment of a catalog that says what exists (according to scientific theories), ontology seems to begin to elude naturalization. As we have argued, another essential part of ontology, as essential as the establishment of that catalog, or inventory, is the establishment of more general ontological categories about which the items in that catalog must be understood.

Ontology qua catalog-ontology can be obtained from science by philosophers of science employing traditional methods of metaontology, such as Quinean strategies for determining ontological commitment: one just needs to identify what are the things that need to exist according to specific formulations of QM. Catalog-ontology, thus, is in a great measure sensitive to formulation. If (Everettian) many-worlds QM uses multiverses as explanatory devices, one may argue that such an approach to QM is ontologically committed (qua catalog-ontology) with many worlds, for example. This is fairly standard in the metaphysics of science, albeit ``ontology'' and ``metaphysics'' are not commonly distinguished on these fronts, generating terminological tensions. Now the second role of ontology, on the other hand, is characterized by the establishment of more general ontological categories, in terms of which all of the existing entities according to the theory must be classified (e.g. ``objects'', ``structures'', ``processes'', etc.); we call this ``type-ontology''. In this sense of ontology qua type-ontology, philosophers of science enter the traditional ontological debates by searching for the most adequate ontological categories to understand the entities obtained by the catalog-ontology.

Methodologically, we argued that only catalog-ontology is naturalizable in the sense of being derived by current science, and that type-ontology is not determined or \textit{completely} epistemically warranted by science. This means that the epistemic virtues one may identify in the defense of a specific type-ontology and its \textit{application in} science are not in any sense obtained from scientific theories whatsoever, and do not derive any kind of direct epistemic warrant from science. Another central difference between the present approach to ontology and current approaches to the metaphysics of science, besides the distinction between two roles for ontology, is to individuate ``ontology'' and ``metaphysics'' by their subject-matter, the latter dealing with nature questions (as opposed to existence questions); so, metaphysics is a step beyond catalog-ontology and type-ontology, whereas current approaches in the metaphysics of science conflate ontology and metaphysics qua metaphysics. By disentangling ontology and metaphysics and, moreover, distinguishing between two kinds of ontology, we can better appreciate the prospects for successfully naturalizing each of them.

In the case of quantum mechanics, theories can indicate the catalog of what exists ($\mathscr{O}_{cat}$), e.g., whether there are particles, worlds, or flashes. However, what theories do not do is indicate whether these same particles, worlds, or flashes, should be understood as the ontological categories of objects, structures, events, processes, and so on. Of course, the list could be extended even further, but the point is that \textit{no} quantum theory indicates this aspect of ontology, which we call $\mathscr{O}_{typ}$. This part of the ontology is developed creatively by ontologists, and, unlike the catalog that is objectively extracted, the $\mathscr{O}_{typ}$ is additional layer of explanation, placed \textit{upon} the $\mathscr{O}_{cat}$, and consequently on the scientific theory in question---and there is nothing completely objective in the association of a $\mathscr{O}_{typ}$ instead of another, since the same $\mathscr{O}_{cat}$ can be understood according to several different $\mathscr{O}_{typ}$.

At the end of the day, the denaturalization of ontology presents us with the following lesson: independence from science comes with the price of leaving behind the promise of an epistemic paradise, whose guarantees were once thought to be granted by naturalization. Ontologists, welcome to freedom: now, by the sweat of your face you will have to gain your ontological categories!

\section*{Acknowledgements}
Order of authorship does not represent priority, as authors contributed equally to this article. Raoni Arroyo acknowledges support from grant \#2022/15992-8, São Paulo Research Foundation (FAPESP). Jonas Arenhart acknowledges partial support from the Brazilian National Research Council (CNPq). Previous versions were presented at the VII Brazilian Society for Analytic Philosophy Conference and at the 13th Principia International Symposium. We would like to thank the audience of both conferences for useful feedback. In particular, to Félix Pinheiro, Gilson Olegario da Silva, Otávio Bueno, and Peter Lewis.

\printbibliography

\end{document}